\documentclass[]{aa}

\usepackage{graphicx}
\usepackage{epsfig}
\usepackage{epstopdf}
\usepackage{txfonts}
\usepackage{natbib}
\usepackage{gensymb}

\newcommand{\raiseChi}

\begin{document} 

\title{WISE J061213.85-303612.5: a new T-dwarf binary candidate\thanks{This work used data collected at the VLT under runs 89.C-0494(A,B,C) and 91.C-0501(A,B).},
   \thanks{Based on observations made with the NASA/ESA Hubble Space Telescope,  and obtained from the Hubble Legacy Archive, which is a collaboration between the Space Telescope Science Institute (STScI/NASA), the Space Telescope European Coordinating Facility (ST-ECF/ESA), and the Canadian Astronomy Data Centre (CADC/NRC/CSA).}}
   \author{N. Hu\'elamo\inst{1}
          \and
          V. D. Ivanov\inst{2,3}
          \and
          R. Kurtev\inst{4,5}
          \and
          J. H. Girard\inst{2}
          \and
          J. Borissova\inst{4,5}
          \and
          D. Mawet\inst{2}
          \and
          K. Mu\v{z}i\'c\inst{2}
          \and
          C. C\'aceres\inst{4}
          \and
          C. H. F.~Melo\inst{2}
          \and 
          M. F.~Sterzik\inst{3}
          \and
          D. Minniti\inst{5,6}
                    }
  \institute{Centro de Astrobiolog\'{\i}a (INTA-CSIC);  ESAC Campus, P.O. Box 78, E-28691 Villanueva de la Ca\~nada, Spain\\
              \email{nhuelamo@cab.inta-csic.es}
         \and
        European Southern Observatory, Alonso de Cordova 3107, Casilla 19, Vitacura, Santiago, Chile               
         \and
         European Southern Observatory, Karl Schwarzschild Strasse 2, Garching bei M\"unchen, D-85748 Germany
         \and
          Instituto de F\'{\i}sica y Astronom\'{\i}a, Universidad de Valpara\'{\i}so, Av. Gran Breta\~na 1111, 2360102, Valpara\'{\i}so, Chile  
          \and 
          Milenium Institute of Astrophysics, MAS
          \and
          Departamento de Ciencias F\'{\i}sicas, Universidad Andr\'es Bello, Rep\'ublica 220, 837-0134, Santiago, Chile
                        }\date{Received December 31, 2014; accepted ----, ----}
  \abstract
   {T and Y-dwarfs are among the coolest and least luminous objects detected, and they can help to understand the properties of giant planets. 
    Up to now, there are more than 350 T dwarfs that have been  identified thanks to large imaging surveys in the infrared, and their multiplicity properties can shed light on the formation process. }
   {The aim of this work is to look for companions around a sample of  seven ultracoool objects. Most of them 
   have been discovered by the WISE observatory and have not been studied before for multiplicity.}
   {We observed a sample six T dwarfs and one 
   L9 dwarf with the Laser Guide Star (LGS) and NAOS-CONICA, the adaptive optics (AO) facility, and the near infrared camera at
   the ESO Very Large Telescope.  We observed  all the objects in one or more near-IR filters ($JHK_s$).}
   {From the seven observed objects, we have identified a subarcsecond binary system,  WISE~J0612-3036, 
   composed of two similar components with spectral types of T6. We measure a separation of $\rho$ = 350$\pm$5\,mas and a position angle of $PA = 235\pm1^{\circ}$. 
   Using the mean absolute magnitudes of T6 dwarfs in the 2MASS $JHK_{\rm s}$ bands, we estimate a distance of $d$=31$\pm$6\,pc and  derive a  projected separation of 
   $\rho\sim11\pm2$\,au. Another target, WISE~J2255-3118, shows a very faint object 
   at 1\farcs3 in the $K_{\rm s}$ image. The object is marginally detected in $H$, and we derive a near infrared color of  $H-K_{\rm s}$$>$ 0.1\,mag.
   $HST/WFC3$ public archival data reveals that the companion candidate is an extended source.Together with the derived color, this suggests that the source is most probably a background galaxy. The five other sources are apparently single, with 3-$\sigma$ sensitivity limits between $H$=19--21 for companions at separations $\geq$ 0\farcs5.}
   {WISE\,0612-3036 is probably a new T-dwarf binary composed of two T6 dwarfs. 
   As in the case of other late T-dwarf binaries, it shows a mass ratio close to 1,  although its projected separation, $\sim$11\,au, is larger than the average ($\sim$ 5\,au).
   Additional observations are needed to confirm  that the system is bound.}
   
   \authorrunning{Hu\'elamo et al.}
   \titlerunning{A new T-dwarf binary system} 
   \maketitle
%

\section{Introduction}

Brown dwarfs (BDs) are the bridge between late-type stars and planetary-mass objects.
T dwarfs (with effective temperatures of $T_{\rm eff}\sim$ 1500-500\,K) and Y-dwarfs ($T_{\rm eff} \leq$ 500\,K ) 
are among the coolest and least luminous objects detected,
and their properties can help us to understand giant extrasolar planets.

In recent years, a significant number of BDs have
been identified in wide-field optical and infrared imaging surveys, e.g.,   
the Two Micron All Sky Survey \citep[2MASS,][]{Skrutskie2006}, the DEep Near Infrared Survey \citep[DENIS,][]{Epchtein1997}, 
or the UKIRT Infrared Deep Sky Survey \citep[UKIDSS,][]{Lawrence2007}.
For the coolest objects, deep near-IR and mid-IR surveys such as the Canada-France Brown Dwarf Survey \citep[CFBDS,][]{Delorme2008}
or the Wide-Field Infrared Survey Explorer \citep[WISE,][]{Wright2010} have contributed  to significantly increasing the  
number of known T and Y-dwarfs \citep[e.g., ][]{Burningham2010,Albert2011,Kirk2011,Cushing2011,Kirk2012,Tinney2012,Pinfield2014,Pinfield2014b}.  
Currently, there are $\sim$355 and 15 identified T  and Y dwarfs, respectively.\footnote{www.dwarfarchives.org}

Multiplicity is directly related to the stellar and substellar formation process,  and it has been 
studied for different  mass regimes. In the case of  T dwarfs, several works have focused on studying the 
companions around these sources \citep[e.g., ][]{Burg2003,Bouy2003,Burg2006,Bouy2008,Gelino2011,Liu2011,Liu2012}.
Most of them have benefited from  laser guide star (LGS) adaptive optics (AO) assisted observations at large 
ground-based telescopes  or/and  the {\em Hubble Space Telescope} (HST) data in order to achieve the best angular resolution and sensitivity.
As a result there are now $\sim$14  identified T-dwarf binaries. Most of them show   
very small separations \citep[$\sim$2-5\,au,][]{Burg2003,Burg2006,Liu2011}, although there are also wide binaries that have been detected 
\citep[e.g.,  WISEJ1711+3500, $\rho=15\pm2$\,au, ][]{Liu2012}. In a recent work,  \citet{aberasturi2014}  have derived a 
binary fraction of  $<$16 or $<$25\%, depending on the  underlying mass ratio distribution.



Motivated by the large number of T and Y dwarfs identified by WISE, we started a program to search for companions to newly
discovered  BDs in 2012 using high angular resolution LGS AO-assisted observations in the near-IR. Most of the selected targets were T and Y dwarfs identified 
with WISE and listed in \citet{Kirk2011}. In this paper we present the results from our AO
imaging campaigns. In Sections~\ref{Samp} and \ref{Observations} we described the sample selection, 
the observations, and the data reduction. In Section~\ref{Results} we  discuss the main results, including the 
sensitivity limits. The main conclusions are summarized in Section~\ref{conclusions}.

\section{The sample of brown dwarfs.\label{Samp}}

\begin{table*}[th!]
\caption{Sample of observed brown dwarfs}\label{sample}
\begin{tabular}{lccllll}\hline
 Target & RA         & DEC      & Sp. Type & $H^*$ & $d$ & Reference\\
             & (J2000) & (J2000) &                & (mag)  & (pc)  & \\ \hline
 WISE J0612-3036 & 06 12 13.8 & -30 36 12.2 & T6 & 17.06$\pm$0.11 & 17.0 & 1\\ 
 2MASS J1114-2618  & 11 14 51.3   & -26 18 23.5 & T7.5 & 15.73$\pm$0.12 & 10$\pm$2 & 2,3 \\ 
 WISE J1436-1814  & 14 36 02.2  & -18 14 21.9 & T8 pec & $>$17.62 & 16.8 & 1,4 \\ 
 WISE J1612-3420   & 16 12 15.9 & -34 20 27.1 & T6.5 & 16.96$\pm$0.03$^{**}$ & 16.2 & 1,4\\ 
 WISE J1959-3338   & 19 59 05.7 & -33 38 33.7 & T8    & 17.18$\pm$0.05$^{**}$ & 11.6 & 1 \\ 
 WISE J2255-3118 & 22 55 40.7 & -31 18 42.0 & T8  & 17.70$\pm$0.11$^{**}$ & 13.0  & 1\\ 
 WISE J2327-2730 & 23 27 28.7 & -27 30 56.5 & L9  & 15.481$\pm$0.103 & 21.7 & 1 \\ \hline
\end{tabular}

{\bf Notes:} $^*$\,2MASS photometry except for $^{**}$ with MKO filter system photometry; $^1$\,\citet{Kirk2011}; 
$^2$\,\citet{Tinney2005}; 
$^3$\,\citet{aberasturi2014}; 
$^4$\,\citet{Kirk2012}
\end{table*}

\begin{table*}
\caption{Observing log.}\label{obslog}
\begin{tabular}{lllllllllr}\hline
Date & Target & Filter & Exp. time & Seeing$^1$ &$t_0$$^1$  & Airmass & LGS mode$^2$ & FWHM$^3$  & Strehl$^4$ \\
 & & & (sec) &   (arcsec) & (ms) & &   & (arcsec) &  (\%) \\ \hline
 
2012-09-08 & WISE 0612-3036  & $J$      & 300 &  0.8  &  2.5 & 1.1  & non-AO & -- & --  \\
                       &                               & $H$     & 180 &  0.7  & 2.5  & 1.3 & $se$ & -- & --   \\  
                       &                               & $K_{\rm s}$   & 420 &  0.7  & 3.0  & 1.2 & non-AO &  -- &  --  \\ 
                       & WISE 2255-3118  & $H$     & 780 & 1.3 & 1.1 & 1.0 & $se$ &  0.50  & 2.1           \\
                       &                               & $K_{\rm s}$ & 840 & 1.0 & 1.5& 1.0 & $se$  &  0.40 & 3.3  \\
                      
                        & WISE 2327-2730 & $H$ & 660 &  0.9 & 1.7& 1.3& $se$  &  0.37 & 3.2 \\
\hline

2013-05-19  & 2MASS J1114-2618          & $J$         &   480 &  0.9& 1.5 & 1.2 & $tt$     & 0.40 & 1.8  \\ 
  &                                                             & $H$       &   480 &  0.9  & 2.0 & 1.1 & $tt$     & 0.22 & 4.8   \\
  &                                                             & $K_{\rm s}$    &   480 &  0.9  & 2.0 & 1.1 & $tt$     &  0.14 & 11.0 \\ 
  & WISE 1959-3338          & $J$        &   600 & 0.9 & 2.0& 1.0     & $tt$ & 0.39 & 2.7\\
    &                                      &  $H$     &   400 & 0.9 & 1.7 & 1.0    & $tt$ & 0.27 & 5.0 \\
   &                                       & $K_{\rm s}$   &   300 & 0.8  & 2.0 & 1.0   & $tt$ & 0.25 & 8.0\\ \hline
2013-07-01 & WISE 1436-1814  & $ H$      & 240  & 1.1  & 4.0 & 1.0  & $se$  & 0.53& 2.3\\  
                       &                               &  $K_{\rm s}$  & 240   & 1.1  & 4.0 & 1.0  & $se$ &  0.29& 4.6\\ 
                       & WISE 1612-3420 & $H$       & 900   & 1.0  & 4.0 & 1.0  & non-AO  & 0.41& 2.4\\ 
                       &                               & $K_{\rm s}$   & 800   & 1.0  & 4.0 & 1.0  & non-AO  & 0.30 & 4.0\\  \hline
\end{tabular}

{\bf Notes:} $^1$ Average values from the Differential Image Motion Monitor (DIMM); $^2$ $se$: seeing enhancer, $tt$: tip-tilt 
(see text for details); $^3$  Full width at half maximum (FWHM) measured on the target; $^4$ Strehl ratio measured on the target, except in the case
of WISE\,0612-3036, which is resolved into a tight binary system.
\end{table*}

Our target list was mainly selected from the newly found BDs with {\it WISE}, 
which is a mid-IR satellite launched in 2009 that surveyed the entire sky in
four bands: 3.4, 4.6, 12, and 22\,$\mu$m.
Finding BDs was a primary science goal of WISE, and the strategy was based on 
a particular color cut: the methane absorption feature near 3.3\,$\mu$m made the [3.6]-[4.6]
color extremely red ($\geq$2\,mag), unique among substellar objects.

\citet{Kirk2011} summarize the first WISE BDs discoveries that
included a list of a hundred of detected substellar objects. 
For our study, we initially selected WISE T and Y dwarfs  from this work. We kept only those
observable from Paranal Observatory and  never studied for multiplicity before
\citep{Gelino2011,Liu2012}. 

Additional dwarfs were discovered and studied afterwards \citep[][]{Cushing2011,Gizis2011,Kirk2012}, and we 
included them in our target list.
Finally, we also considered several bright targets (mainly previously known L-type and early-T dwarfs) as backups to be  observed under bad conditions.
One of  these targets, 2MASS J1114-2618, was observed by the $HST$ in the meantime, and the
results were recently presented by \citet{aberasturi2014}. We include it here for comparison.
In total, we selected $\sim$50 targets for our study.



\section{Observations and data reduction \label{Observations}}

The observations presented here were obtained with NAOS-CONICA (NACO),
the AO system and near-IR camera  \citep{Lenzen2003,Rousset2003} at the
Very Large Telescope (VLT). Since most of the targets are too faint to be used as
natural guide stars (NGS), we made use of the laser guide star (LGS)
facility at the VLT.  

To obtain a full AO correction, the LGS facility should be used in combination with
a tip-tilt ($tt$) reference star brighter than $V$\,=\,17\,mag, located at a maximum separation of 40$\arcsec$ from the target.
In the absence of a $tt$ star, the LGS can be used in the so-called `seeing
enhancer' ($se$) mode \citep{Girard2010}, that is, LGS-assisted AO observation but without correcting for tip/tilt. 
Because the low orders of the wavefront corrugations are not corrected, the SE 
mode provides an image quality intermediate between simple (non-AO) imaging  
and full LGS closed loop observations. (We refer to the NACO manual for a full description
of this mode. \footnote{http://www.eso.org/sci/facilities/paranal/instruments/naco/doc)} 
For our observations, we used a $tt$ reference star whenever available,
and switched to $se$ mode otherwise.

The observations were allocated in seven different nights over 2012 and 2013.  
We could not collect any data in four of the nights
(2012-06-10, 2012-07-12, 2012-12-08, and 2013-03-07)  because of inadequate atmospheric
conditions for AO and the LGS and/or technical issues associated to the laser itself. 
The atmospheric conditions in the three other nights were moderate to poor for the LGS,
with average coherence times ($t_0$) $\le$ 3\,ms and variable seeing. 
In these conditions we observed a total of 20 targets, but we could only obtain good quality datasets 
for seven of them. They are included in Table~\ref{sample}. As shown in the table, 
some of the targets were even observed without AO owing to the unsuitable atmospheric conditions and 
poor LGS performance.

All the targets were observed with the S54 objective, providing a field of
view of 54$\arcsec$$\times$54$\arcsec$ and a nominal plate scale of
0\farcs054. The data was obtained in one or several near-IR filters
($JHK_s$), depending on the atmospheric conditions and
brightness. To estimate and remove the sky background, we observed the targets
applying random offsets between the individual exposures (`Autojitter' mode).

The data was reduced using {\em eclipse} \citep{Devi1997} and
following the standard steps for near-IR imaging data: the images were dark-subtracted and
flat-field-divided,  and bad pixels were masked out. The sky background was estimated 
and subtracted from all the individual images. Finally, the frames were shifted 
and co-added to generate the final stacked image. Since the atmospheric conditions 
were variable during the observations,  some of the  individual frames showed a low image quality.
To improve the final spatial resolution of the combined data, we
removed the worst  individual images, i.e. those with the lowest Strehl ratio and the largest FWHM, and kept only those with the best AO performance.

The observing log  is given in Table~\ref{obslog}.  As can be seen, the seeing and coherence time were
not optimal for AO+LGS observations, and this is reflected in the final FWHM and Strehl ratios measured on the combined images.

\section{Results~\label{Results}}

\begin{figure}[!t]
\includegraphics[angle=0,scale=0.5]{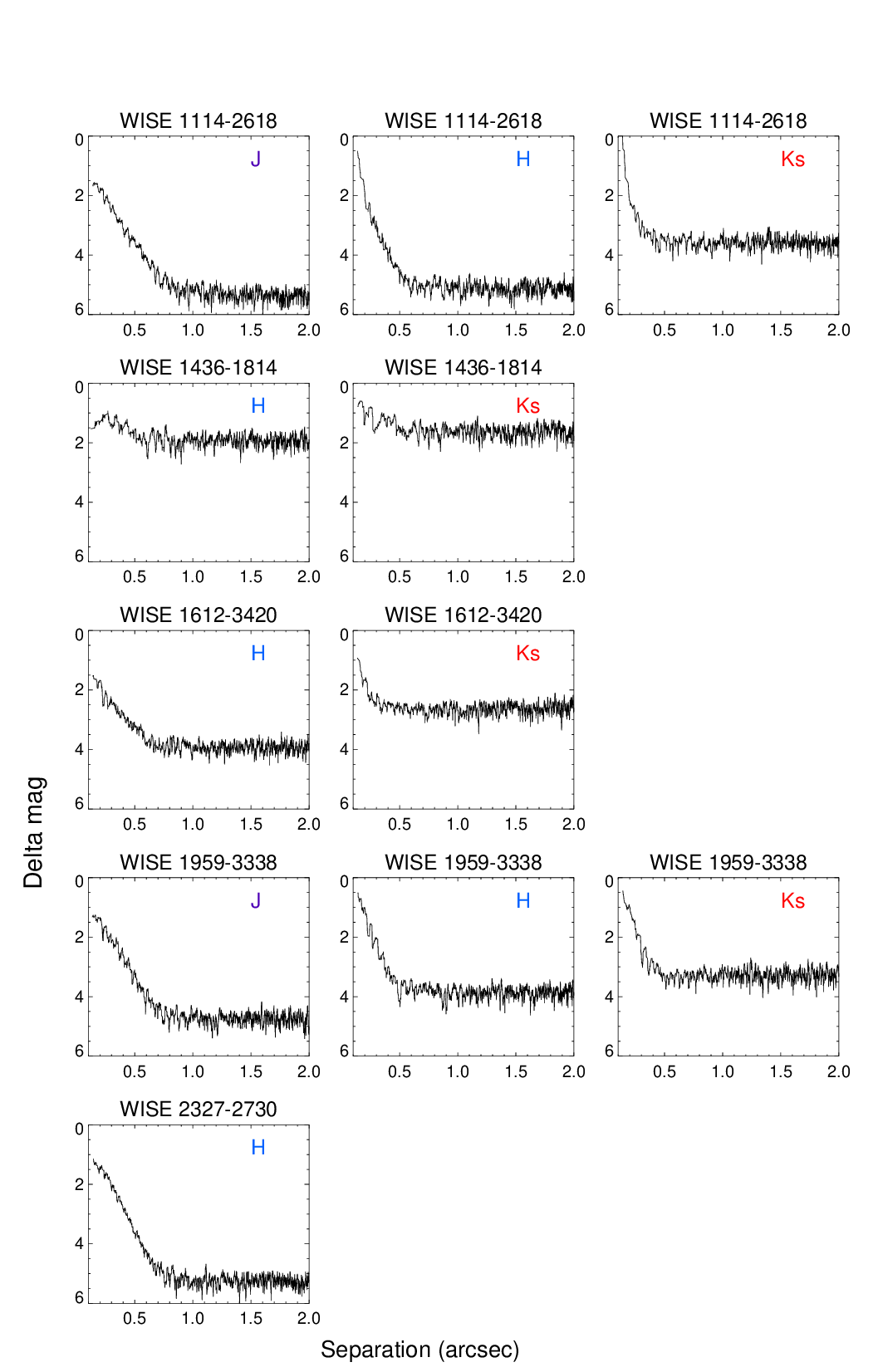}
\caption{Sensitivity limit of the NACO observations. We show the curves for the 
five apparently single sources, one target per row, in the different
NACO filters. They were computed as the  3-$\sigma$ standard deviation of the
PSF radial profile. }\label{sensit}
\end{figure}

Five of the seven BDs observed with NACO+LGS are apparently single.
We estimated the sensitivity limit as a function of the separation  for all of them (see Fig.~\ref{sensit}) 
in the different NACO filters.
The curves have been estimated from the point spread function (PSF) radial profiles of
the targets, while computing the standard deviation of the 
noise in bins of 20 pixels and one pixel
step.  We derived the difference in magnitudes at a
3-$\sigma$ level normalizing by the peak intensity of the PSF.

Usually, the best combination of sensitivity and angular resolution is achieved in the H band.
To derive the sensitivity limits in this band, we used the $H$ mag listed in Table~\ref{sample}.
For the targets with MKO filter system photometry, we converted the magnitudes into the 
2MASS system using the relations from \citet{Stephens2004}. We obtain $H_{\rm 2MASS}\sim H_{\rm MKO}-0.04$\,mag for
the given spectral types. We reach 3-$\sigma$ magnitude limits between  $H$$\sim$19-21\,mag for separations of  0\farcs5 (see Table~\ref{limits}).
For an average distance of $\sim$15\,pc and ages of 1\,Gyr and 5\,Gyr, this magnitude interval corresponds to masses of  $\sim$11-7\,$M_{\rm Jup}$, 
and $\sim$30-18\,$M_{\rm Jup}$, respectively,  according to COND 
evolutionary tracks \citep{Chabrier2000}.
For each  individual target, we estimated the approximate spectral type (Sp. Type $_{\rm lim}$) that corresponds to the 
$H$-band limiting magnitude  (3-$\sigma$) using the  $M_{\rm H}$ versus spectral type relation derived by \citet{Kirk2012}, and the
individual distances from Table~\ref{sample}. This value is included 
in the last column of Table~\ref{limits}.
In the case of WISE\,1114-2618, we achieve a $\Delta H$ value of 4.7\,mag at 0\farcs5. This is deeper than the HST/F164N observations presented by \citet{aberasturi2014}, with a contrast of 1.5\,mag at a separation of 0\farcs6.

\begin{table}\label{limits}
\caption{Contrast and sensitivity limits (3-$\sigma$) in the $H$-band at a separation of 0\farcs5. The magnitudes are provided in the 2MASS system.}
\begin{tabular}{llll}
\hline
Target & $\Delta H$ & $H_{\rm lim}$@0.5\arcsec & Sp. type $_{\rm lim}$$^1$ \\ 
            &  (mag) & (mag) & \\ \hline
2MASS\,1114- 2618 & 4.7 & 20.4 &  Y0\\
WISE\,1436-1814     &  2.0 & $>$19.6 & $>$T9\\
WISE\,1612-3420     & 3.2 & 20.2 &  T9\\
WISE\,1959-3338     & 3.5 & 20.7  &  Y0\\
WISE\,2327-2730     & 3.7 & 19.2  &  T8\\ \hline
 \end{tabular}
 
 $^1$ Spectral type that corresponds to the limiting $H$-mag according to the $M_{H}$-Sp.~Type relation from
 \citet{Kirk2012}.
 
\end{table}

\subsection{WISE\,0612-3036}

\begin{figure}[!t]
\includegraphics[angle=0,scale=0.54]{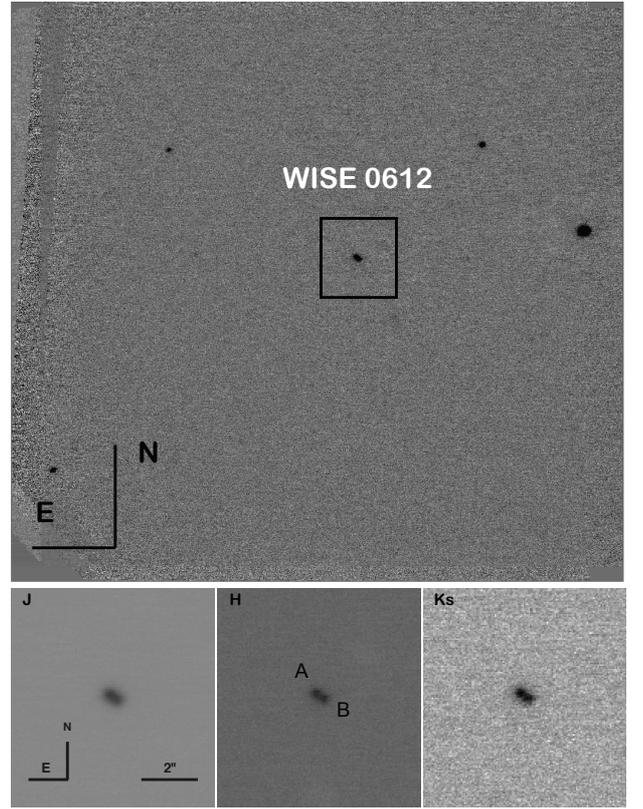}
\caption{NACO/LGS observations of WISE\,0612-3036. The top panel shows the $\sim$54\arcsec$\times$54\arcsec
NACO FOV. The target, WISE-0612-3036, is in the center of the black square. The star 
2MASSJ06121227-3036100, used to calibrate the JHKs photometry of the binary system, is the brightest 
source in the west. The bottom panels show a zoom on the newly discovered binary in $JHK_s$.}\label{fig1}
\end{figure}

 We have resolved WISE 0612-3036 (WISE0612, hereafter)  into a  subarcsecond binary 
(Fig.~\ref{fig1}). The two objects are detected in the three bands, $JHK_s$
with an estimated separation of  $\rho$\,=\,350$\pm$5\,mas and 
position angle of PA\,=\,235$\pm$1$^{\circ}$. The values are derived computing the average 
and standard deviation of the  measurements in the three bands. We note that only the H-band data was 
obtained with LGS assisted AO,  while the subsequent $JK_{\rm s}$ 
data was obtained in open loop owing to technical problems with the laser.

We performed PSF stellar photometry of the binary system with ALLSTAR in
DAOPHOT~II \citep{Stetson1992}. For the photometry we used the final stacked images in each
filter. The PSF was constructed using three or four stars in the field of view for the $J$ and $HK_{\rm s}$ bands, respectively. The typical PSF
errors of both components are in the range of 0.012--0.034\,mag. For the
transformation to the standard 2MASS $JHK_{\rm s}$ system, we used the
only star in the field with 2MASS photometry: 2MASS\,J06121227-3036100
with $\alpha$(2000)=6$^{\rm h}$ 06$^{\rm m}$ 12\fs27 and $\delta(2000)=-30\degr 36\arcmin 10\farcs0$ and 
apparent magnitudes of $J$=15.915$\pm$0.080\,mag, $H$=15.141$\pm$0.066\,mag, and $K_{\rm s}$=15.024$\pm$0.139\,mag. 
The final errors of the photometry are weighted means of the PSF errors and errors of the standard star.  
To double check the PSF photometric calibration with the bright star  
2MASS\,J06121227-3036100, we estimated its $K_{\rm s}$-band magnitude using a standard 
star observed at the end of the night.  We find that the difference between the 2MASS tabulated value 
and our calibrated photometry is only 0.04\,mag.

The estimated magnitudes for the two components in the 2MASS system are 
provided in Table~\ref{bin0612}. \citet{Kirk2011} report magnitudes of  $J$=17.00$\pm$0.09, $H$=17.06$\pm$0.11, and 
$K_{\rm s}$= 17.34$\pm$0.21 for the unresolved system. If we correct them for binarity, the 
difference between our PSF photometry and these values are -0.12, 0.14, and 0.16 in $JHK_{\rm s}$, respectively, which are consistent
within the given uncertainties. 

The  primary (A)  is the brightest object located to the NE  (see Fig.~\ref{fig1}). 
\citet{Kirk2011} classify the unresolved object as a T6 dwarf based on near-IR spectroscopy.
The difference in magnitudes between the two components are $\Delta J$= --0.05$\pm$0.14\,mag, 
$\Delta H$= --0.10$\pm$0.14\,mag, and $\Delta K_{\rm s}$=--0.13$\pm$0.24\,mag, so we can assume similar spectral types for the two objects.

\begin{figure}[!t]
\includegraphics[angle=0,scale=0.55]{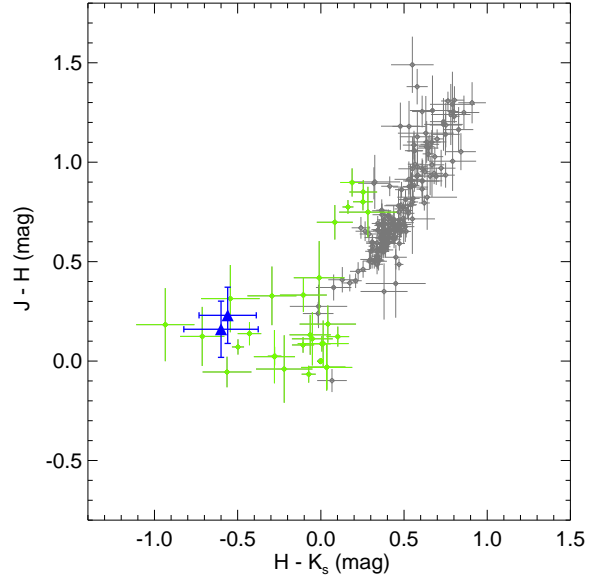}
\caption{Color-color near-IR diagram. We have included a sample of ultracool dwarfs from \citet{Dupuy2012}: 
the small gray diamonds represent M- and L-dwarfs, while the
green diamonds are  T dwarfs. We have only plotted  objects with photometric uncertainties 
smaller or equal to 0.2\,mag. The binary components of WISE0612 are represented by blue triangles.}\label{ccd}
\end{figure}

We have plotted the binary components in a near-IR color-color diagram (see Fig.~\ref{ccd}) and compared them  
with well known T-type dwarfs  from \citet[][Table 10]{Dupuy2012} with available 2MASS photometry. We have only
included objects  with photometric uncertainties smaller than 0.2\,mag in the three bands. The WISE0612 system falls 
within the region of the late-T dwarfs with  a very blue  $H-K_{\rm s}$ color. As explained in \citet{Burg2002}, this is expected in late-T dwarfs since 
the 2.1$\mu$m peak  starts to flatten until it is suppressed for the latest type objects. This is indeed the case of WISE0612, which shows 
no  emission peak at 2.1 $\mu$m in the near-IR spectrum obtained by \citet{Kirk2011}.

\begin{table}
\caption{Properties of the binary system WISE\,0612-3036\,AB.}\label{bin0612}
\begin{tabular}{lrr}
 & NACO photometry & \\ \hline 
Parameter    & WISE 0612A & WISE 0612B \\ \hline
$J$            & 17.85$\pm$0.10  & 17.89$\pm$0.10 \\
$H$          & 17.62$\pm$0.10    & 17.73$\pm$0.10 \\
$K_{\rm s}$       & 18.18$\pm$0.14    & 18.33$\pm$0.20\\ \hline 
$J-H$                       &     0.25$\pm$0.14      &    0.20$\pm$0.14            \\
$H-K_{\rm s}$                  &  --0.60$\pm$0.17    & --0.60$\pm$0.22  \\ 
$J-K_{\rm s}$                    & --0.40$\pm$0.17   &  --0.42$\pm$0.22 \\ \hline
                                 & Binary parameters & \\ \hline
$\rho$ (mas)           & 350$\pm$ 5&  \\
$\rho$ (au)             & 11$\pm$2   &  \\
PA (deg)                   & 235$\pm$1   & \\
Distance (pc)           & 31$\pm$6   & \\ \hline
\end{tabular}
\end{table}

The spectrophotometric distance to the unresolved source, based on its WISE W2 magnitude and 
its spectral type,  is $\sim$17\,pc \citep{Kirk2011}.
To derive the distance to the new binary system we have 
the mean 2MASS absolute magnitudes as a function of the spectral type from \citet{Dupuy2012}. 
The values derived for a T6 dwarf are
$M_{J}$=15.35$\pm$0.02, M$_{H}$=15.32$\pm$0.02\,mag, and $M_{K_{\rm s}}$=15.68$\pm$0.01\,mag, with typical dispersions of 
0.11, 0.17, and 0.33\,mag, respectively. Using these values, we derived distances of $\sim$ 31$\pm$2\,pc ($J$-band), 29$\pm$3\,pc ($H$-band) and 32$\pm$5\,pc ($K_{\rm s}$-band) and adopted an average  value of $d$ = 31$\pm$6\,pc for the rest of the paper.

\begin{figure}[!t]
\includegraphics[angle=0,scale=0.46]{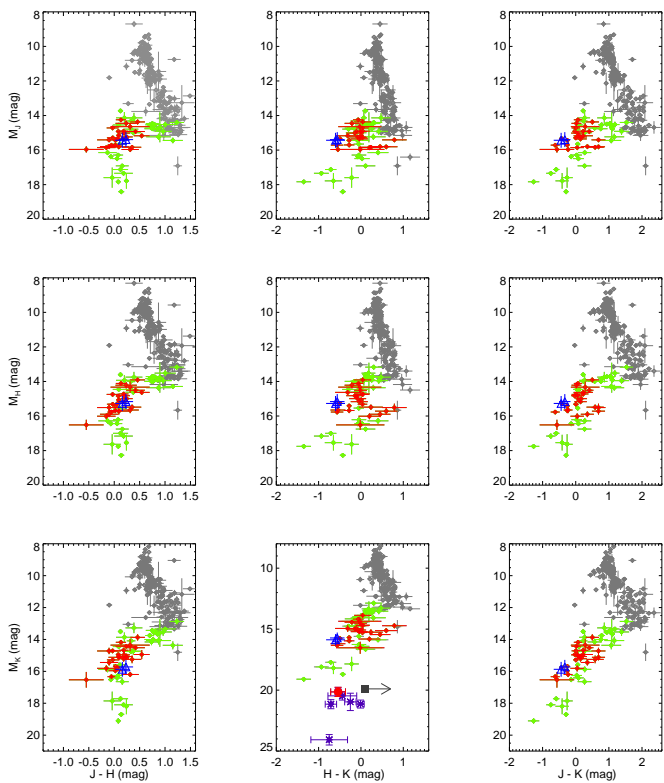}
\caption{Near-IR color-magnitude diagrams. We represent a sample of ultracool dwarfs with measured parallaxes and 2MASS photometry from 
\citet{Dupuy2012}:  gray diamonds represent ML-dwarfs  and green diamonds the T dwarfs. 
We have highlighted in red the T-dwarfs with spectral types between T5 and T7. The binary components of WISE0612 are represented 
by blue open triangles. In the lower central panel, the faint companion candidate to WISE2255 is represented by a filled square and
WISE\,1711+3500B \citep[T9.5,][]{Liu2012}  by a filled circle. A sample
of 4 Y-dwarfs \citep{Liu2012,Leggett2013} are indicated by asterisks.}\label{cmd}
\end{figure}

Assuming that the two objects are bound, we represent the
binary components in several color-absolute magnitude diagrams (see
Fig.~\ref{cmd}), together with a sample of MLT dwarfs with the measured
parallaxes included in \citet{Dupuy2012}.  We excluded objects
with 2MASS photometric errors larger than 0.5\,mag, and relative
errors in their parallaxes ($\delta\pi/\pi$) larger than 0.6.  As
expected, the two binary components of WISE0612 lie in the region
between T5-T7 dwarfs, with very blue $H-K_{\rm s}$ and $J-K_{\rm s}$
colors.

For the assumed distance of 31$\pm$6\,pc,  the projected separation of the binary is  $\rho \sim$ 11$\pm$2\,au. Since the age of the system is unknown, we estimated the mass of the objects assuming two ages of  $\sim$ 1\,Gyr and 5\,Gyr.
According to COND evolutionary tracks \citep[2MASS system,][]{Chabrier2000}, we derived  
masses of  $\sim$30\,$M_{\rm Jup}$ (1\,Gyr) and  $\sim$60\,$M_{\rm Jup}$ (5\,Gyr) for the individual components. For a circular orbit, we derived orbital periods  of $\sim$150\,yr for 1\,Gyr and $\sim$105\,yr for 5\,Gyr.

Finally, we note that a single-epoch observation is not enough to  classify the binary as bound.
Although additional data is needed to confirm it as a common proper motion pair, the small angular separation 
makes a random coincidence unlikely.   
We have estimated the  probability of a background object lying close to the target using the number
sources detected in the NACO $H$-band image. We detect eight sources in the $\sim$56\arcsec$\times$56\arcsec field of view. 
Since the companion candidate is detected at a separation of 0\farcs35,  the probability of detecting an object
 in an area of $\pi$$\times$$0\farcs35^2$= 0.3848 arcsec$^2$ around the target is  8$\times$(0.3848/56$^2$), that is, 0.09\%. 
 If we only consider the detected sources with a similar brightness to the companion (2), we derive a probability of 0.02\%.
 We have also checked that there is no object at the given position in the Palomar Observatory Sky Survey (POSS) II images.
 
 On the other hand, the proper motion of the primary estimated by
 \citet{Kirk2011} is $\mu_\alpha.cos\delta$ = -137$\pm$22\,mas/yr and
 $\mu_\delta=$-256$\pm$21\,mas/yr, that is, $\sim$0.3\arcsec/yr.
 Therefore, the binary can be confirmed with second-epoch non-AO
 observations obtained after 3-4\,yrs, to allow the BD enough time to
 move far enough from the companion, if it turns out that the latter
 is not co-moving.

\subsection{WISE 2255-3118}

\begin{figure}[!t]
\centering
\includegraphics[angle=0,scale=0.45]{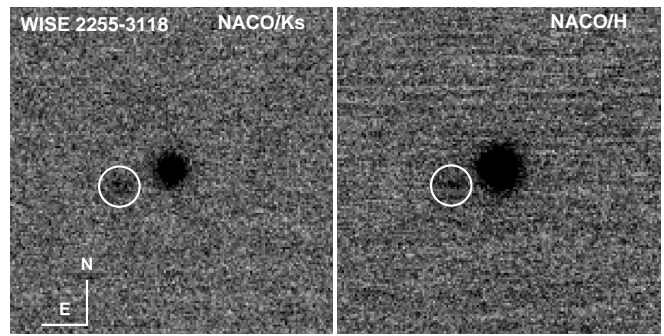}
\caption{NACO/LGS $K_{\rm s} \& H$ observations of WISE\,2255-3118. We display  a field of 8\arcsec$\times$8\arcsec. A faint source (encircled) is detected at $\sim$1\farcs3 SE from the T dwarf.}\label{fig2}
\end{figure}

In the case of the T8 dwarf WISE\,J2255-3118 (WISE2255, hereafter),
the final image quality was not good enough to detect very close
companions. However, we have detected a very faint source in the
$K_{\rm s}$ band at a separation of $\sim$1\farcs3 (see
Fig.~\ref{fig2}). This source is marginally detected in the $H$ band.
The estimated distance to the source, 13\,pc \citep{Kirk2011}, implies
a projected separation of $\sim$17\,au.

We performed aperture photometry on the $HK_s$-band images and calibrated it using standard stars observed during the night. We
derive $H$=17.64$\pm$0.18\,mag and $K_{\rm s}$= 17.47$\pm$0.25\,mag
for the primary.  For the faint object we
estimate $K_{\rm s}$=20.5$\pm$0.3\,mag and $H >$20.6\,mag and a color of
$H-K_{\rm s}$ > 0.1\,mag.
 
 If bound, the companion candidate would be at a distance of 13\,pc, which implies a $M_{\rm Ks}$$\sim$19.9\,mag. When we compare its properties
 to those of the T dwarfs represented in the  central lower panel of Fig.~\ref{cmd}, we see that the object does not follow the sequence of the
 coldest T dwarfs, displaying a redder $H-K_{\rm s}$ color.
 In this panel we also compared  WISE2255 with WISE\,1711+3500, a physical binary system with very similar properties:
 it consists of a T8 primary  and a T9.5 companion with $\Delta K$=3\,mag
 \citep{Liu2012}. The difference in the $H$ band between the two components is $\Delta H$ of -2.83, and  its
  $H-K_{\rm s}$$_{\rm , MKO}$ color is -0.42$\pm$0.17. Using the relations by \citet{Stephens2004} for a T9 dwarf (the coldest object in their sample),
 we can convert its magnitudes into the 2MASS system ($K_{\rm s,2MASS}$ = $K_{\rm s,MKO}$ +0.16, and $H_{\rm 2MASS}$ = $H_{\rm MKO}$ +0.04), obtaining $H-K{\rm s}$$_{\rm , 2MASS}$ = -0.54, significantly bluer than the upper limit reported for the
 WISE2255 companion candidate.

 Finally, we also compared the photometry of the companion candidate with five Y dwarfs included in \citet{Liu2012} and \citet{Leggett2013}.
 Since they all have $HK_s$ photometry in the MKO system, we have converted them to the  2MASS system following the same procedure described above.
 The result is displayed in Fig.~\ref{cmd} (low central panel). The companion is redder than the Y dwarfs, but it lies close enough to their locus,  preventing us from drawing a firm conclusion about the nature of this object.

\begin{figure}[!t]
\centering
\includegraphics[angle=0,scale=0.44]{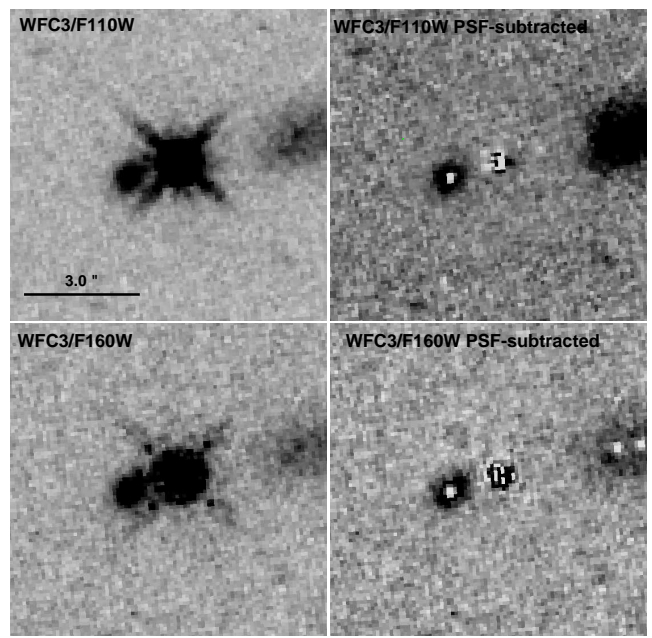}
\caption{HST/WFC3 F110W and F160W reduced images of WISE\,2255-3118. We display a field of 8\arcsec$\times$8\arcsec. North is up and east to the left. The companion candidate is clearly resolved in the two filters. The PSF-subtracted images are displayed on the right panels, where the 
companion candidate shows large residuals consistent with an extended source.}\label{hst}
\end{figure}

 We searched in public archives for additional data on this source and
 found $HST$/WFC3 data obtained in two near-IR filters, F110W and
 F160W (Program 12972, P.I. Gelino). The observations were obtained on
 2012 October 17 in MULTIACCUM mode and following a four dithering
 pattern. The total exposure time was $\sim$1200\,sec in the two
 filters. We retrieved the final calibrated images ({\em flt} name
 extension) from the Hubble Legacy Archive (HLA) and reprocessed them
 using the routines within the  DrizzlePac\footnote{http://drizzlepac.stsci.edu/} package. In
 particular, we used {\sc multidrizzle} to combine the exposures into
 a master image, resampling onto a finer pixel scale and correcting
 for geometric distortions in the camera. (We selected a final\_scale
 of 0.09 "/pixel and final\_pixfrac of 0.8.)
The T dwarf  and the companion are clearly detected in the final $HST$ images (see Fig.~\ref{hst}), and the analysis of the two images with {\sc SEXtractor} \citep{Bertin1996} and {\sc PSFEx} \citep{Bertin2011}  reveals that the faint companion candidate is an extended source. To visualize it, we represented the HST PSF-subtracted images in Figure~\ref{hst}, where the companion shows strong residuals in comparison to the T dwarf. This is consistent with the concentration index (CI) derived by the $HST$ team and included in the HLA {\sc SEXtractor} catalog for the F160W band (Data Release 8). This index is defined as the difference in magnitude from two different apertures (0\farcs09 and 0\farcs27), and it is used to separate extended from point-like sources. They measure a CI\,=1.4  for the companion candidate, while point-like sources in the WFC3 show 0.5 $<$ CI $<$ 1.0.
  We note that the $HST$ image is relatively crowded, with a large number of spatially resolved galaxies. In fact, the HLA catalog contains 
  275 detected sources, with 235 displaying CI $>$ 1.0.

  An extended source is consistent with a background galaxy or with a binary. In the latter case, and if bound to the primary, 
  the binary components  would be cooler than T8, and probably in the Y-dwarf regime.
  However, with the current data we cannot distinguish between the two scenarios.
      
  That the detected source is extended, together with the red $H-K_{\rm s}$ color, suggests that the faint object is most probably
  a background galaxy, although additional imaging is needed to test whether the objects are co-moving or not.
  


\begin{table*}
\caption{Summary of known T-dwarf binary systems}\label{allbinT}
{\tiny
\begin{tabular}{lllllll} \hline
Name & SpT A & SpT B & Distance    & $\rho$ & $\rho$  & Reference \\
           &                     &       & [pc]             &      [mas]   &    [au]  & \\ \hline

SDSS\,102109.69-030420.1            & T1     & T4       &  29$\pm$4       & 172$\pm$5  & 5.0$\pm$0.7 & \citet{Burg2006} \\ 
2MASS J14044941-3159329         & T1 &T5       &   $\sim$23       & 133.6$\pm$0.6 & $\sim$3.1         & \citet{Looper2008} \\ 
$\epsilon$ Indi~B                               & T1     &  T6     & 3.626$\pm$0.013 & 732$\pm$2 & 2.654$\pm$0.012& \citet{McCaughrean2004} \\ 
SDSS\,153417.05+161546.1            & T1.5  &T5.5 &  $\sim$36       & 110$\pm$5  & $\sim$4         &   \citet{Liu2006} \\ 
SIMP J1619275+031350AB               & T2.5 & T4     &    22$\pm$3  &    691$\pm$2             & 15.4$\pm$2.1 & \citet{Artigau2011} \\
SDSS\,092615.38+584720.9           & T4     & T4       & 38$\pm$7        & 70$\pm$6    & 2.6$\pm$0.5  &  \citet{Burg2006}\\ 
WISEPA\,184124.73+700038.0       & T5     & T5       & 40.2$\pm$4.9  & 70$\pm$14 & 2.8$\pm$0.7  & \citet{Gelino2011} \\ 
2MASS J15344984-2952274         & T5    & T5.5    &  16$\pm$5       & 140.3$\pm$0.57  & 2.3$\pm$0.5 & \citet{Burg2003,Liu2008} \\ 
2MASS J12255432-2739466         & T6 &T8       &  11.2$\pm$0.5 & 282$\pm$5       &  3.17$\pm$0.14     & \citet{Burg2003} \\ 
2MASS J15530228+1532369       & T6.5 & T7    &  12$\pm$2    & 349$\pm$5  & 4.2$\pm$0.7          & \citet{Burg2006} \\ 
WISEPA\,J171104.60+35000036.8 & T8  &T9.5    &  19$\pm$3        &  780$\pm$2 &  15.0$\pm$2.0     & \citet{Liu2012} \\ 
WISEPA\,J045853.90+643452.6   & T8.5 & T9    &  10.5$\pm$1.4& 510$\pm$20  & 5.0$\pm$0.4        & \citet{Gelino2011} \\ 
WISE J014656.66+423410.0         & T9    & Y0      & 10.6$\pm$1.5  & 87.5$\pm$2.0  & 0.93$\pm$0.14  & \citet{Dupuy2015} \\ 
WISEPC\,J121756.91+162640.2   & T9    & Y0       &  10.5$\pm$1.7  & 759.2$\pm$3.3  & 8.0$\pm$1.3 &  \citet{Liu2012}\\
CFBDSIR J1458+1013                         & T9.5 &T10  &  23.1$\pm$2.4  & 110$\pm$5 & 2.6$\pm$0.3 & \citet{Liu2011} \\ 
\hline
\end{tabular}
}
\end{table*}

\subsection{Comparison with T-dwarf binaries}

\begin{figure}[!t]
\includegraphics[scale=0.56]{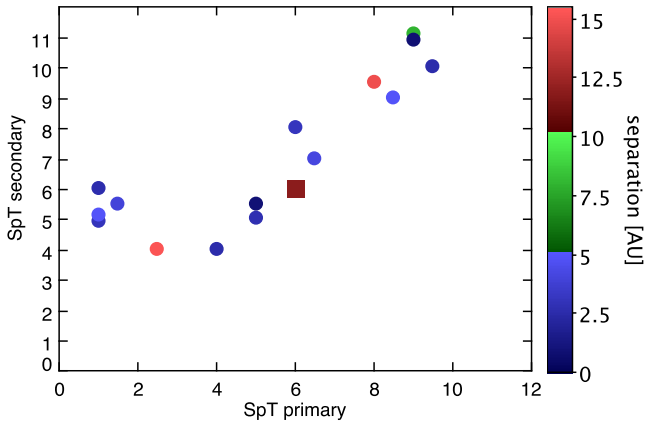}
\caption{Known T-dwarf binaries. We have represented the spectral types of the secondaries versus the primaries for known T-dwarf binaries (circles). 
0 stands for T0, 10 for T10, and 11 for Y0.
The color code is related with the projected separation (in au) of the binaries. Most of them show separations smaller than 5\,au, while only four
show $\rho \ge$ 8\,au. WISE\,0612-3036 (filled square) is among these four binaries with a projected separation of 11\,au.}\label{tbinaries}
\end{figure}

To put our study into context, we compared the properties of  WISE\,0612-3036 with other T-dwarf binaries from
different AO and HST studies (see Table~\ref{allbinT}). We have represented the spectral types of the secondaries 
versus the spectral types of the primaries  in Figure~\ref{tbinaries}.  Their projected separation in the sky (in au)
have been color-coded. As seen in Figure~\ref{tbinaries}, all the early-T primaries with SpT $<$ T3  have  secondaries later
or equal than T4, while most of the late-T primaries (SpT $\geq$ T4) show nearly equal mass companions 
(typical errors in spectral typing are between 0.5-1 type). The separations range between $\sim$2-16\,au, with 
only four binaries showing separations larger than or equal to 8\,au. WISE\,0612-3036 is among these four binaries with a projected 
separation of 11\,au.

\section{Conclusions}\label{conclusions}

We  have observed seven BDs with the NACO+LGS system to look for close companions.
In general, the observing conditions were not optimal for the LGS, resulting in poor
corrections that did not allow the innermost regions around the targets to be explored.

Our main findings can be summarized as follows:

\begin{itemize}

\item Five targets are apparently single with 3-$\sigma$ limiting magnitudes of  $H$ between 19--21 mag at 0\farcs5.

\item The T-dwarf  WISE\,0612-3036 is clearly resolved into a subarsecond binary. PSF-photometry of the binary 
has allowed us to derive the individual $JHK_{\rm s}$ magnitudes of the two components, which  are consistent with a very 
similar spectral type of T6. Using absolute magnitude-spectral type  relationships in $JHKs$, we estimate a distance  
to the binary of  $d\sim31\pm6$\,pc, which implies a projected separation 
of  $\rho=11\pm2$\,au. Additional observations are needed to confirm the system as bound.

\item The target WISE\,2255-3031  shows a very faint source at  a separation of 
1\farcs3 SE from the primary.  Its  $H-K_{\rm s}$ color ($>$\,0.1\,mag) is redder than the colors of the latest T dwarfs and Y dwarfs. 
Since $HST$ public archival data in two near-IR filters shows that the source is extended, 
we suggest  that it is most probably a  background galaxy.

\item{Similar to other known late T-dwarf binaries, WISE\,0612-3036 is comprised of two  nearly equal-mass objects.  
Its projected separation, 11\,au, is larger than the average value for T dwarf binaries ($\sim$5\,au).}

\end{itemize}

\begin{acknowledgements}
We thank the referee, C. Reyle,  for her useful comments. This research has been funded by Spanish grants AYA2010-21161-C02-02 and AYA2012-38897-C02-01. Support for NH, JB,RK, and DM is provided by the Ministry of Economy, Development, and Tourism's Millennium Science Initiative 
through grant IC120009,  awarded to The Millennium Institute of Astrophysics, MAS. JB is supported by FONDECYT No.1120601, 
RK is supported  by Fondecyt Reg. No. 1130140.  DM is also supported by FONDECYT No.1130196 and by the Center for Astrophysics 
and Associated Technologies CATA PFB-06. CC acknowledges the support from project CONICYT FONDECYT Postdoctorado 3140592.
We are indebted to the Paranal staff for their support during the runs. NH thanks H. Bouy for useful discussions. This research has benefited from the M, L, T, 
and Y dwarf compendium housed at DwarfArchives.org. 
\end{acknowledgements}

\bibliographystyle{aa}
\bibliography{wisebd}

\end{document}